\documentclass[12pt]{article}

\usepackage{amsmath,amsfonts,bbold}
\usepackage{amssymb}
\usepackage[usenames,dvips]{color}
\usepackage{graphicx}
\usepackage{booktabs}
\usepackage{macros}
\newcommand{\dmat}[3]{{\begin{pmatrix}#1\\&#2\\&&#3 \end{pmatrix}}}

\usepackage{fullpage}

\makeatletter
\@addtoreset{equation}{section}
\makeatother

\newtheorem{theorem}{Theorem}

\begin{document}

\title{Universal Approximations for Flavor Models}
\author{Gero von Gersdorff\\
\normalsize Pontif\'icia Universidade Cat\'olica, Rio de Janeiro, Brazil}
\date{}
\maketitle

\begin{abstract}
We develop a systematic analytical approximation scheme for the singular value decompositions of  arbitrary complex three dimensional matrices $Y$ with non-degenerate singular values. We derive exact expressions for the errors of this approximation and show that they  are bounded from above by very simple ratios of the form $(y_i/y_j)^{2n}$ where $y_i<y_j$ are singular values of $Y$ and $n$ is the order of the approximation. The applications we have in mind are the analytical and numerical treatments of arbitrary theories of flavor. We also compute upper bounds for  the errors of the Cabbibo Kobayashi Maskawa (CKM) matrix that only depend on the ratios of the masses and the physical CKM angles.
\end{abstract}

\section{Introduction}

One of the unresolved mysteries of the Standard Model (SM) is the peculiar structure of the fermion sector, in particular the very non-generic structure of their masses and mixings.
Disregarding the neutrino sector, the properties of the fermions are encoded in their Yukawa couplings, complex three by three matrices $Y_u$, $Y_d$ and $Y_e$.

\begin{table}
\centering
\begin{tabular}{@{}cc c cc c cc c|c cc c cc@{}}
\toprule
$y_u$	& $6.3\times 10^{-6}$	&&$y_d$	& $1.4 \times 10^{-5}$	&&$\theta_{12}$	&$0.23$			&&&
	$y_e$	&$2.8 \times 10^{-6}$	&&$\theta_{12}$&0.58\\
$y_c$	& $3.1\times 10^{-3}$	&& $y_s$	& $2.7 \times 10^{-4}$ 	&& $\theta_{23}$	& $4.2\times 10^{-2}$&&&
	$y_\mu$	&$6.0 \times 10^{-4}$&&$\theta_{23}$	&0.82\\
$y_t$	& $0.87$				&& $y_b$	& $1.4 \times 10^{-2}$	&& $\theta_{13}$	& $3.7\times 10^{-3}$&&&	
	$y_\tau$	& $1.0 \times 10^{-2}$&&$\theta_{13}$&0.15\\
\bottomrule
\end{tabular}
\caption{Quark and Lepton data at 1 TeV in the SM \cite{Antusch:2013jca}.}\label{tab:2}
\end{table}

A brief summary of the SM fermion data is given in Tab.~\ref{tab:2}. The actual values depend on the renormalization group scale (and scheme) as well as on possible New Physics thresholds (such as supersymmetry), we quote the SM values in the $\overline {\rm MS}$ scheme at 1 TeV as given in Ref.~\cite{Antusch:2013jca}.
Notice that RG running and threshold corrections typically give only $\mathcal O(1)$ modifications to these numbers.
One observes the  hierarchical structure
\be
y_{u_1}\ll y_{u_2}\ll y_{u_3}\,,\qquad
y_{d_1}\ll y_{d_2}\ll y_{d_3}\,,\qquad
y_{e_1}\ll y_{e_2}\ll y_{e_3}\,.
\label{eq:hiery}
\ee
Moreover, the CKM mixing angles follow the hierarchy
\be
\theta_{13}\ll \theta_{23}\ll \theta_{12}\ll 1 \,.
\label{eq:hierV}
\ee
while mixing angles in the neutrino sector are $\mathcal O(1)$.

The observations Eq.~(\ref{eq:hiery}) require that the eigenvalues of the Hermitian matrices $Y_xY_x^\dagger$ ($x=u,d,e$) are very hierarchical, while Eq.~(\ref{eq:hierV}) implies that the eigenvectors of $Y_u Y_u^\dagger$ and $Y_d Y_d^\dagger$ are closely aligned. 
Clearly, the latter fact means that the up and down Yukawa couplings have to "know of each other" in some way. 

Many models have been proposed to explain this peculiar structure, we comment on a few representative ones in Sec.~\ref{sec:ex}.
In order to have good analytical and numerical control over the model parameter space, it is common practice to expand eigenvalues and eigenvectors in terms of some small parameters present in a given model. However the goodness of such expansions not only depends on the size of the expansion parameters (of which there might be several) but also on all the other parameters of the model. To judge its accuracy, one would have to  go higher order in the expansion, and in order to achieve a given precision one has to resum the expansion up to a certain order.
In this short paper we comment on a very powerful and fully model-independent approximation scheme that can be carried to arbitrarily high precision. It is extremely simple to apply and does not rely on the existence of any expansion parameter. The expressions for eigenvalues and mixings are given exclusively in terms of the $(Y_x)_{ij}$, without any assumptions on their sizes. The errors of the approximation are exactly bounded by (not just of the order of) simple ratios of the quantities $(y_i/y_j)^{2n}$, $y_i<y_ j$, where $n$ is the order of the approximation. 
We will see that known approximations of particular models follow without any calculation, with the added bonus of adding complete analytical control over their errors.

We hope that the method presented in this paper can help to better understand the parameter space of existing and yet to be conceived models of flavor, facilitating for instance fits to the data. Finally this paper may serve as a reference for students studying mainstream models (such as Frogatt-Nielsen \cite{Froggatt:1978nt}) for the first time who want a simple and easy to follow derivation of the known approximations.

\section{Preliminaries}
\label{sec:prelim}

We recall that an arbitrary complex matrix $X$ can be written as a so-called singular value decomposition (SVD)
\be
X=U_{L}\mathcal X U_R^\dagger\,,
\ee
where $U_L$ and $U_R$ are unitary and $\mathcal X$ is diagonal. 
The elements of $\mathcal X$ are called singular values (SV) of $X$, which are unique up to phases. 
Given a particular phase convention for $\mathcal X$ (say all SVs real positive), the remaining ambiguity consists of multiplication of $U_L$ and $U_R$ with the same diagonal phases matrix from the right. The columns of $U_L$ ($U_R$) are eigenvectors of $XX^\dagger$ ($X^\dagger X$), and $|\mathcal X|^2$ are the eigenvalues.

Our discussion will be greatly simplified by a convenient parametrization of a general unitary matrix $U\in U(3)$ which we write as
\be
U=K(a,b,c)\begin{pmatrix}e^{i\alpha}\\& e^{i\beta}\\&& e^{i\gamma}\end{pmatrix}\,,
\label{eq:param}
\ee
with $K$ defined as 
\be
K(a,b,c)\equiv\begin{pmatrix}
\frac{1}{n_1}& \frac{a-c^*(b-ac)}{n_1n_3}   &\frac{b}{n_3}\\
\frac{-a^*}{n_1}&\frac{1+b^*(b-ac)}{n_1n_3}&\frac{c}{n_3}\\
\frac{-(b-ac)^*}{n_1}&\frac{-c^*-ab^*}{n_1n_3}&\frac{1}{n_3}\\
\end{pmatrix}\,.
\label{eq:abc}
\ee
We will refer to this parametrization as the $abc$ parametrization.
The quantities $n_1$ and $n_3$ are normalization constants for the column vectors, given by
\be
n_1\equiv\sqrt{1+|a|^2+|b-ac|^2}\,,\qquad n_3\equiv\sqrt{1+|b|^2+|c|^2}\,.
\label{eq:n1n3}
\ee
Notice that $\det K=1$.
The numbers $a,b,c$ are arbitrary complex numbers, which together with the phases $\alpha,\, \beta,\, \gamma$ comprise the 9 (real) degrees of freedom of $U(3)$. We will sometimes use the shorthand 
\be
\tilde b\equiv b-ac.
\ee
Given an arbitrary $U\in U(3)$, one can readily compute these parameters:
\be
a=-\frac{U^*_{21}}{U^*_{11}}\,,\qquad \tilde b=-\frac{U^*_{31}}{U^*_{11}}\,,\qquad b=\frac{U_{13}}{U_{33}}\,,\qquad c=\frac{U_{23}}{U_{33}}\,.
\ee 
and
\be
\alpha=\arg U_{11}\,,\qquad \gamma=\arg {U_{33}}\,,\qquad \alpha+\beta+\gamma=\arg\det U\,.
\ee

Unitary Matrices that appear in SVDs of random matrices (with matrix elements drawn from a uniform measure) are  distributed according to the invariant Haar measure \cite{Haba:2000be}. 
In the  parametrization Eq.~(\ref{eq:param}), the Haar measure of the group is simply 
\be
dU=\frac{1}{32\pi^6}\frac{1}{n_1^4n_3^4}da\,da^*db\,db^*dc\,dc^*d\alpha\, d\beta\, d\gamma\,,
\label{eq:haar}
\ee
where we normalized the measure to unity.

\section{Approximate Diagonalizations}
\label{sec:svd}

In this section we are going to describe a neat analytic way of diagonalizing a Hermitian matrix.  Even though in principle it can be applied to any matrix, it is particularly suited for matrices that have hierarchical spectra, such as the mass matrices of the SM.

Let $A$ be a positive-definite, Hermitian 3 by 3 matrix, and let $a_i$ be its (nonndegenerate) eigenvalues. By convention, we will assume that $a_1<a_2<a_3$. 

Let us consider the matrix $A^n$ which of course has the same eigenvectors as $A$. Diagonalizing it as
\be
(A^n)_{ij}=a^n_k U_{ik}U^*_{jk}\,,
\label{eq:An}
\ee
we write the three column vectors of $A^n$ suggestively as
\bea
(A^n)_{i1}&=&a_3^nU_{13}^*\left(
U_{i3} +\left[\frac{a_2^n}{a_3^n}\frac{U_{12}^*}{U_{13}^*}\right] U_{i2}+
\left[\frac{a_1^n}{a_3^n}\frac{U^*_{11}}{U_{13}^*}\right] U_{i1}
\right)\,,
\\
(A^n)_{i2}&=&a_3^nU_{23}^*\left(
U_{i3}+\left[\frac{a_2^n}{a_3^n} \frac{U_{22}^*}{U_{23}^*}\right]U_{i2}+\left[\frac{a_1^n}{a_3^n}\frac{U^*_{21}}{U_{23}^*}\right] U_{i1}
\right) \,,
\\
(A^n)_{i3}&=&a_3^nU_{33}^*\left(
U_{i3}+\left[\frac{a_2^n}{a_3^n} \frac{U_{32}^*}{U_{33}^*}\right]U_{i2}+\left[\frac{a_1^n}{a_3^n}\frac{U^*_{31}}{U_{33}^*}\right] U_{i1} 
\right)\,.
\eea
Observe that we have written each column of $A^n$ as a linear combination of the eigenvectors of $A$.
This way it becomes clear that as $n\to \infty$ each of the three columns of $A^n$ (when properly normalized) converges to $U_{i3}$, the eigenvector corresponding to the largest eigenvalue $a_3$, and hence we can  compute the $b$ and $c$ parameters of $U=K(a,b,c)$ as
\be
b=\lim_{n\to \infty}\frac{(A^n)_{1j}}{(A^n)_{3j}}\,,\qquad c=\lim_{n\to \infty}\frac{(A^n)_{2j}}{(A^n)_{3j}}\,.
\ee

We can choose any of the three representations $j=1,2,3$, but observe that for the $j$th column of $A^n$ a small $|U_{j3}|$ will slow down the convergence (or even destroy it if $U_{j3}=0$) because it appears in the denominator of the subleading terms. However the $U_{j3}$ cannot all be small simultaneously, in fact unitarity implies $\max_j |U_{j3}|\geq \frac{1}{\sqrt 3}$. Fortunately it is possible to find 
the $j$ that maximizes $|U_{j3}|$ without actually knowing $U$, just by looking at the length of the columns
\be
\sum_i |(A^n)_{ij}|^2
=|U_{j3}|^2a^{2n}_3+|U_{j2}|^2a^{2n}_2+|U_{j1}|^2a^{2n}_1\,.
\ee
It is easy to see that in the limit of $n\to \infty$ the largest $|U_{j3}|$ is exactly correlated with the longest column. For finite $n$ one can still find a lower bound for $|U_{j3}|$ which is evaluated in App.~\ref{sec:error} and comes very close to $\frac{1}{\sqrt 3}$.  We thus choose the longest of the columns of $A^n$ to achieve the smallest error.
 
For the parameter $a$ and $\tilde b$ which are related to the smallest eigenvalue, we proceed with the same method, this time with the matrix $A^{-n}$,
\be
a^*=-\lim_{n\to \infty}\frac{(A^{-n})_{2k}}{(A^{-n})_{1k}}\,,\qquad
\tilde b^*=-\lim_{n\to \infty}\frac{(A^{-n})_{3k}}{(A^{-n})_{1k}}\,,
\ee
where for completeness we also give the dependent quantity $\tilde b$.
Notice that if the columns used to determine $U_{i3}$ and $U_{i1}$ are different, $j\neq k$, the eigenvectors are exactly orthogonal (this trivially follows from the fact that $A^{-n}$ and $A^n$ are inverses) and equivalently $\tilde b=b-ac$. On the contrary, it is possible that the maximum length criterion gives the same column (say the first one of both $A^n$ and $A^{-n}$), in this case they are only orthogonal up to small corrections (see App.~\ref{sec:error} for details).
Sometimes it might be more convenient to work with the matrix of minors
\be
\tilde A\equiv \det (A) (A^{-1})^T\,,
\ee
in which case one has
\be
a=-\lim_{n\to \infty}\frac{(\tilde A^{n})_{2k}}{(\tilde A^{n})_{1k}}\,,\qquad
\tilde b=-\lim_{n\to \infty}\frac{(\tilde A^{n})_{3k}}{(\tilde A^{n})_{1k}}\,.
\ee
This way, the procedure is defined also for $a_1=0$.

Obviously, these results  hold for matrices of any dimensions $d$, but the advantage for the case of interest of $d=3$ is that 
it already completely determines the full unitary matrix $U$, as
the eigenvector to $a_2$ follows from 
\be
U_{i2}^{(n)}=\eps_{i\ell m}U_{\ell 3}^{*(n)}U_{m1}^{*(n)}\,,
\ee
or, equivalently, from the $abc$ representation in Eq.~(\ref{eq:abc}).

Having found an approximation $U^{(n)}$ for the matrix $U$, our next goal is to find an upper bound for  the error of our algorithm. To this end, let us define the matrix
\be
V\equiv U^\dagger U^{(n)}\,,
\label{eq:defV}
\ee
which must converge to the identity. An explicit expression for $V$ in terms of $U_{ij}$ and the $a_i$ can easily be obtained and is given in App.~\ref{sec:error}, where it is also shown that
$V$ can be bounded by 
\be
\begin{pmatrix}
1-|V_{11}|^2	&	|V_{12}|^2	&	|V_{13}|^2\\
|V_{21}|^2		&	1-|V_{22}|^2&	|V_{23}|^2\\
|V_{31}|^2		&	|V_{32}|^2	&	1-|V_{33}|^2
\end{pmatrix}
\leq 2\begin{pmatrix}
(\frac{a_1}{a_2})^{2n}	& 	(\frac{a_1}{a_2})^{2n}	&	(\frac{a_1}{a_3})^{2n}\\
(\frac{a_1}{a_2})^{2n}	&	(\frac{a_1}{a_2})^{2n}+(\frac{a_2}{a_3})^{2n}	&	(\frac{a_2}{a_3})^{2n}\\
(\frac{a_1}{a_3})^{2n}	&	(\frac{a_2}{a_3})^{2n}	&	(\frac{a_2}{a_3})^{2n}
\end{pmatrix}\,.
\label{eq:errors}
\ee
Eq.~(\ref{eq:errors}) is valid for $j\neq k$ (which is commonly satisfied in many explicit models) while for the case $j=k$ a modified prescription yields the estimate Eq.~(\ref{eq:bounds2}).

We summarize our findings in the following theorem.
\begin{theorem}
The longest column of the matrix $A^n$ ($A^{-n}$) converges to the  eigenvector to the largest (smallest) eigenvalue of $A$. The errors are encoded in the matrix V defined in Eqns.~(\ref{eq:defV}) and can be bounded by Eqns.~(\ref{eq:errors}) or (\ref{eq:bounds2}) respectively.
\end{theorem}

This procedure works extremely well and can provide us with very useful analytic expressions and even fast converging numerical ones without the need for the calculation of the eigenvalues.\footnote{Notice that for high-precision numerical calculations one can conveniently work with $n=2^k$ which only requires $k\sim\log n $ matrix products instead of $n$.}

\begin{figure}
\begin{center}
\includegraphics[width=7.5cm]{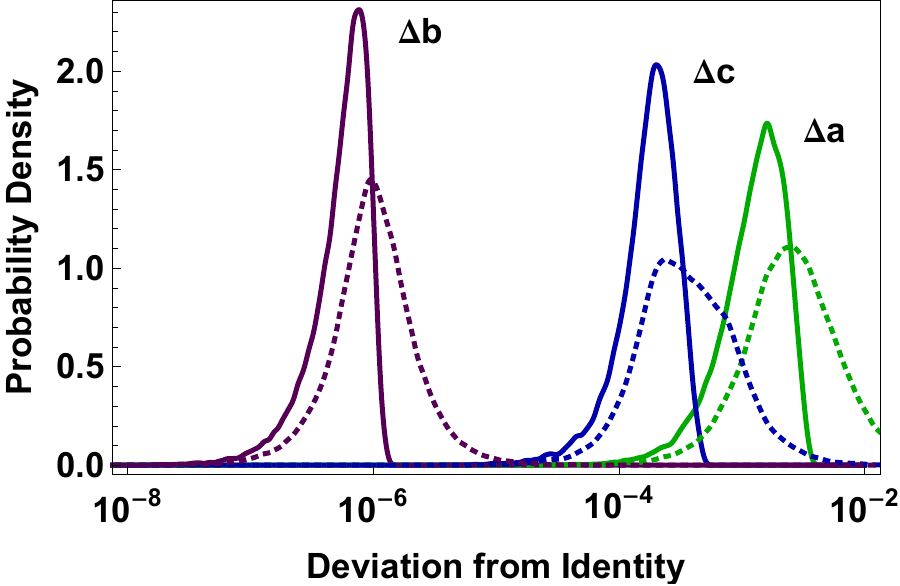}
\end{center}
\caption{Left: error of the approximate eigenvectors for down-type Yukawa couplings in arbitrary bases (see text for details). The dotted lines are the corresponding approximations without the maximum length criterion (instead choosing say $j=1$ always), showing considerably worse errors.}
\label{fig:diag}
\end{figure}

To illustrate the importance of the maximum-length criterion, let us numerically examine the error of our method for a hierarchical matrix with and without the maximum-length criterion. For definiteness, we consider $A=Y_dY_d^\dagger$ with $Y_d$ the down quark Yukawa couplings (see Tab.~\ref{tab:2}) in an arbitrary basis. Starting with the diagonal matrix, we perform random $\mathcal O(1)$ rotations $U$ (drawn from the distribution Eq.~(\ref{eq:haar})) and then recompute the matrix $U\approx U^{(n)}$ using our method with $n=1$. 
To measure the error, we compute the matrix $V\equiv U^\dagger U^{(n)}$ and find its corresponding parameters $\Delta a$, $\Delta b$, and $\Delta c$ which indicate the deviation of $V$ from the identity  and are plotted in Fig.~\ref{fig:diag}.
The errors are  in accordance with the upper limits derived above, while dropping the maximum-length criterion results in considerably larger corrections.

The eigenvalues themselves can be conveniently obtained by computing the traces of $A^n$ and $A^{-n}$. 
\be
a_3=\lim_{n\to \infty} (\tr A^n)^\frac{1}{n}\,,\qquad  a_1=\lim_{n\to \infty}(\tr A^{-n})^{-\frac{1}{n}}\,,
\label{eq:ev1}
\ee
with $a_2$ following from the determinant. Alternatively, one can compute the limits
\be
a_3=\lim_{n\to \infty} \frac{(A^{n+1})_{jj}}{(A^n)_{jj}}\,,\qquad
a_1=\lim_{n\to \infty} \frac{(A^{-(n+1)})_{kk}}{(A^{-n})_{kk}}\,,
\label{eq:ev2}
\ee
where $j$ and $k$ are determined from the maximum length criterion.

Even though this procedure works for any Hermitian matrix (one can even extend the formalism to the case of degenerate spectra), it is particularly well-suited 
for very hierarchical matrices with  spectra $a_1\ll a_2\ll a_3$, such as the mass matrices of the SM. In these cases one obtains very good approximations from the above relations even with $n=1$. 
Notice that we do not need to assume a particular form for the matrices, as long as the spectrum is hierarchical. These approximations thus go beyond the ones usually quoted in the literature, such as those resulting from the particular structure in Eq.~(\ref{eq:FN}), which can be obtained as special cases of the above method.

If $A$ is given as $A=YY^\dagger$, one can also obtain "half-integer" approximations, which can easily be found by starting from the SVD of $Y$, $YY^\dagger Y$, etc, instead of  Eq.~(\ref{eq:An}). 
Of particular interest is the $n=\frac{1}{2}$ case,  obtained by starting from the SVD of $Y$, and results in the following important special case
\begin{theorem}
($n=\frac{1}{2}$ approximation.)
The singular value decomposition of a complex matrix $Y$ is approximately determined as follows. The columns $(U_L)_{i3}$ and $(U_L)_{i1}$ are proportional to the longest columns of the matrices $Y$ and $(Y^{-1})^\dagger$ respectively, and the columns $(U_R)_{i3}$ and $(U_R)_{i1}$ are proportional to  the longest  columns of the matrices $Y^\dagger$ and $ Y^{-1}$.
The errors are then bounded by  Eqns.~(\ref{eq:errors}) or (\ref{eq:bounds2}) with $(a_i)^n\to y_i$.
\end{theorem}
As before, instead of $Y^{-1}$ one might equivalently work with the matrix of minors $\tilde Y=\det Y (Y^{-1})^T$.

We will verify in Sec.~\ref{sec:ex} that for the special structure Eq.~(\ref{eq:FN}) the case $n=\frac{1}{2}$ immediately yields the known approximations. Some corrections (such as the ones reported in Ref.~\cite{Cabrer:2011qb}) can be recovered from the $n=1$ approximation. Finally we notice that there exist approximate expressions for the eigenvalues for half-integer $n$. In the case of $n=\frac{1}{2}$, they read
\be
a_3=\frac{(YY^\dagger)_{jj}(Y^\dagger Y)_{j'j'}}{|Y_{jj'}|^2}\,,\qquad
a_1=\frac{(\tilde Y\tilde Y^\dagger)_{kk}(\tilde Y^\dagger \tilde Y)_{k'k'}}{|\tilde Y_{kk'}|^2}\,,\qquad
\ee
where here $j$ ($j'$) labels the longest row (column) of $Y$ and analogously for $\tilde Y$.

Let us now move to the CKM matrix $V_{\rm CKM}\equiv U_u^\dagger U_d$. The latter reads in our approximation:
\be
V_{\rm CKM}^{(n)}= (U_u^{(n)})^\dagger U_d^{(n)}=V_u^\dagger  V_{\rm CKM}V_d\,.
\ee
From the last expression, by use of the triangle inequality and  Eq.~(\ref{eq:errors}), one can easily obtain
upper bounds on the errors:
\footnote{We only consider the case $j\neq k$ for simplicity. We have used $|(V_{\rm CKM})_{ij}|\leq \theta_{ij}$ and for clarity omitted strictly subleading terms in the upper bounds.}
\bea
|\Delta V_{\rm CKM}|_{12}&\leq& \eps^d_{12} +\eps^u_{12} + \theta_{13} \eps^d_{23}\,,\\
|\Delta V_{\rm CKM}|_{23}&\leq& \eps^d_{23} + \eps^u_{23} + \theta_{13} \eps^u_{12}\,, \\
|\Delta V_{\rm CKM}|_{13}&\leq& \eps^d_{13} + \eps^u_{13} + \theta_{12} \eps^d_{23}   + \theta_{23} \eps^u_{12}  +\eps^d_{23}  \,,\eps^u_{12} 
\eea
where
\be
\eps^u_{ij}\equiv \sqrt{2}\left(\frac{y_{u_i}}{y_{u_j}}\right)^{2n_u}\,, \qquad
\eps^d_{ij}\equiv \sqrt{2}\left(\frac{y_{d_i}}{y_{d_j}}\right)^{2n_d}\,.
\ee
Using the explicit values given in Tab.~\ref{tab:2}, we notice that for any model of Yukawa couplings, the approximation with $n_d=1$ and $n_u=\frac{1}{2}$ gives already excellent accuracy (at most 5\% with the exception of $\theta_{23}$ that has a maximal error of 12\%, dominated by $\eps^u_{23}$). Observe that our approximation is  very well-suited for numerical fits. Away from the physical values for the Yukawas it might lead to large errors (when for instance the true eigenvalues are not hierarchical), nevertheless near the $\chi^2$ minimum one can always fully trust it.

We stress once more that the upper bounds for the errors are very conservative, and can be much less in particular models. 
This typically happens when the entries in the matrix Y are itself hierarchical. One can get a good idea of this effect by computing $V$ approximately using the approximate result for $U$. The result is the matrix $V^{(n)}$, given in Eq.~(\ref{eq:Vn}).

Let us also comment on the PMNS matrix for neutrinos. If neutrinos are moderately hierarchical, one can obtain decent approximations for low $n$. This is only possible for the normal hierarchy, as in the inverted case one has $\frac{m_1}{m_2}>0.98$. On the other hand, in the normal ordering case, one has $0.17\leq \frac{m_2}{m_3} \leq 1$ and $0\leq \frac{m_1}{m_2}\leq 1$ depending on the size of $m_1$. 
One can achieve accuracy comparable with (or less than) the current experimental one at $n=1$ if $m_1\lesssim  10^{-3}$ eV.

\section{Examples}

\label{sec:ex}

In this section, we would like to illustrate our method in three particular models or classes of models.

\subsection{Frogatt-Nielsen and similar}

Let us first turn to the Froggatt-Nielsen Model \cite{Froggatt:1978nt}. 
Identical reasoning applies to models with similar structure as in Eq.~(\ref{eq:FN}), such as extra dimensions \cite{Gherghetta:2000qt,Huber:2000ie} or certain Clockwork models \cite{Alonso:2018bcg}.
The structure of the Yukawa couplings is given as follows
\be
Y_{u}=\dmat{\eps^{q_1}}{\eps^{q_2}}{\eps^{q_3}}\hat Y_{u}\dmat{\eps^{u_1}}{\eps^{u_2}}{\eps^{u_3}}\,,\qquad
Y_{d}=\dmat{\eps^{q_1}}{\eps^{q_2}}{\eps^{q_3}}\hat Y_{d}\dmat{\eps^{d_1}}{\eps^{d_2}}{\eps^{d_3}}\,.
\label{eq:FN}
\ee
where $\hat Y_{u,d}$ are $\mathcal O(1)$ complex matrices and the $q_i$, $u_i$ and $d_i$ are the Froggatt-Nielsen charges of the doublet quarks, up-quarks, and down quarks  respectively, taken to be positive, and $\epsilon$ is a moderately small order parameter. For definiteness, we will consider  the charge assignments \cite{Babu:2009fd}
\be
q=(4,2,0)\,,\qquad u=(4,2,0)\,,\qquad d=(2,1,1)\,,
\ee
with $\eps\approx 0.2$.
In the $n=\frac{1}{2}$ approximation, this selects
the third row of $Y_u^*$ and the first row of $\tilde Y_u$  determine $U_R^u$
\be
(U^u_R)_{i3}\propto Y^{u*}_{3i}\,,\qquad (U^u_R)_{i1}\propto \tilde Y^{u}_{1i}\,,
\ee 
and the third row of $Y_d^*$ and the first row of $ \tilde Y_d$  determine $U_R^d$:
\be
(U^d_R)_{i3}\propto Y^{d*}_{3i}\,,\qquad (U^d_R)_{i1}\propto \tilde Y^{d}_{1i}\,.
\ee
Similarly, it selects the third column of $Y_u$ and the first column of $\tilde Y_u^*$ to determine $U^u_L$,
\be
(U^u_L)_{i3}\propto Y^u_{i3}\,,\qquad (U^u_L)_{i1}\propto \tilde Y^{u*}_{i1}\,.
\ee
For $U_L^d$ we have to choose the first column of $\tilde Y_d$, and either the second or third column of $Y_d$
\be
(U^d_L)_{i3}\propto   \left\{\begin{matrix}
Y^d_{i2}&|\hat Y^d_{32}|>|\hat Y_{33}|\\
Y^d_{i3}&|\hat Y^d_{32}|<|\hat Y_{33}|
\end{matrix}\right.    \,,\qquad (U^u_L)_{i1}\propto \tilde Y^{u*}_{i1}\,.
\ee
One sees that one can directly read off the eigenvectors from the Yukawa matrices and its inverses. For the determination of the third column of $U_L^d$, it is instructive to  compare with the $n=1$ approximation, which is always given by the third column of $Y^dY^{d\dagger}$, or
\be
(U^d_L)_{i3}\propto  \hat Y^{d*}_{33}\,Y^d_{i3}+\hat Y^{d*}_{32}\,Y^d_{i2}+\eps \hat Y^{d*}_{31}\,Y^d_{i1}\,,
\ee
which is essentially a weighted average of the two $n=\frac{1}{2}$ cases, plus an $\epsilon$-suppressed admixture of the first column (which by itself is already numerically suppressed compared to the second and third).
The results here coincide with the ones that one would obtain when making a careful (but rather lengthy) expansion in terms of $\eps$ (see for instance Ref.~\cite{Babu:2009fd}).  However, no calculation is ever necessary to obtain them, one simply reads them off from our standard rules.
Notice that $U^{u,d}_L$ and $U_R^u$ are almost diagonal while $U_R^d$ is almost block-diagonal. This implies that the four error matrices $V_{L,R}^{u,d}$ are even more diagonal than the conservative upper bound in Eq.~(\ref{eq:errors}).

\subsection{Textures from spontaneously broken $SU(2)$ symmetry}
As a second example, we chose the particular texture
\be
Y_{13}=Y_{31}=Y_{11}=0\,,\qquad Y_{12}=-Y_{21}\,,
\label{eq:texture}
\ee 
which has been explored  originally in \cite{Barbieri:1995uv,Barbieri:1997tu} (See Ref.~\cite{Dudas:2013pja} for a variant taking into account more recent measurements of the CKM angles).
We will implement this texture in the following completely general parametrization:
\be
Y=\begin{pmatrix}
0 & \eta \sqrt{y_1y_2} &0\\
-\eta \sqrt{y_1 y_2}	& \eta^2(\beta+\alpha^2 )y_2	&\alpha \rho^{-1} \sqrt{y_2y_3}\\
0	&\alpha\rho\sqrt{y_2y_3}	&\eta^{-2}y_3
\end{pmatrix}\,,
\ee
which gives the inverse
\be
(Y^\dagger)^{-1}=\frac{1}{y_1y_2y_3}\tilde Y^*=\begin{pmatrix}
\frac{\beta^*}{y_1}	&\frac{1}{\eta}\frac{1}{\sqrt{y_1y_2}}& -\eta\alpha\rho \frac{1}{\sqrt{y_1y_3}}\\
-\frac{1}{\eta}\frac{1}{\sqrt{y_1y_2}}	&0	&0\\
\eta\alpha\rho^{-1}\frac{1}{\sqrt{y_1y_3}}	&0	&\eta^2 \frac{1}{y_3}
\end{pmatrix}\,.
\ee
Using redefinitions of the fermion fields, we have removed all the phases except one, chosen to be  $\arg\beta$. 
We stress that this parametrization is exact and no approximations have been made so far. It depends on five real parameters, $\eta$, 
$\rho$ $\alpha$, $|\beta|$, and $\arg \beta$.
Notice that the determinant constraint $\det Y=y_1 y_2 y_3$ is already implemented, but the constraints from the traces
\be
\tr YY^\dagger=y_3^2+y_2^2+y_1^2\approx y_3^2\,,\qquad
\tr (YY^\dagger)^{-1}=y_1^{-2}+y_2^{-2}+y_3^{-2}\approx y_1^{-2}  \,,
\label{eq:traces}
\ee
will lead to two more relations between the five parameters that we will work out below.\footnote{The original matrix, defined by the constraint Eq.~(\ref{eq:texture}), had ten free parameters subject to four possible phase redefinitions. After implementing the determinant and trace constraints, we will essentially have traded three of the six irreducible parameters by the eigenvalues $y_i$, leaving over three free parameters, taken to be $\rho$, $\arg\beta$ and $\eta$.}
However, one can already obtain various useful inequalities. First notice that the traces of $YY^\dagger$ and its inverse imply $|Y_{ij}|\leq y_3$ and $|(Y^{-1})_{ij}|\leq y_1^{-1}$, in particular one has 
\be
1\leq \eta \leq\sqrt{\frac{y_3}{y_1}}\,,
\label{eq:eta}
\ee
\be
|\beta|\leq 1\,,\qquad  |\alpha^2+\beta|\leq\frac{1}{\eta^2}\frac{y_3}{y_2}\,,
\ee
\be
\alpha\rho^{\pm 1}\leq \sqrt{\frac{y_3}{y_2}}
\,,\qquad
\alpha\rho^{\pm 1}\leq \frac{1}{\eta}\sqrt{\frac{y_3}{y_1}}\,,
\ee 
from which one can for instance see that $|Y_{21}|^2$, $|Y_{12}|^2$, $|(Y^{-1})_{21}|^2$ and $|(Y^{-1})_{12}|^2$  contribute only very little to the traces in Eqn.~(\ref{eq:traces}).

The parameter $\eta$ interpolates between various qualitatively different regimes. We will be focusing on the lower end of the interval Eq.~(\ref{eq:eta}),  more precisely
\be
1\leq \eta^2 \ll \frac{y_3}{y_1}
\,,
\label{eq:eta2}
\ee
which provides the most interesting phenomenological models. Under this assumption one finds immediately that Eqns.~(\ref{eq:traces})  reduce to
\bea
|\beta|&=&1\,,
\label{eq:betasol}
\\
\alpha^2&=&\left(\sqrt{\sinh^2 \xi+\eta^4}-\cosh\xi\right)\eta^{-4}\frac{y_3}{y_2}\,.
\label{eq:alphasol}
\eea
Besides $\arg \beta$
this leaves as the only free parameters  $\eta$ (constrained by Eq.~(\ref{eq:eta2})) and $\xi$ or $\rho$ (unconstrained).

\begin{figure}
\centering
\includegraphics[width=7cm]{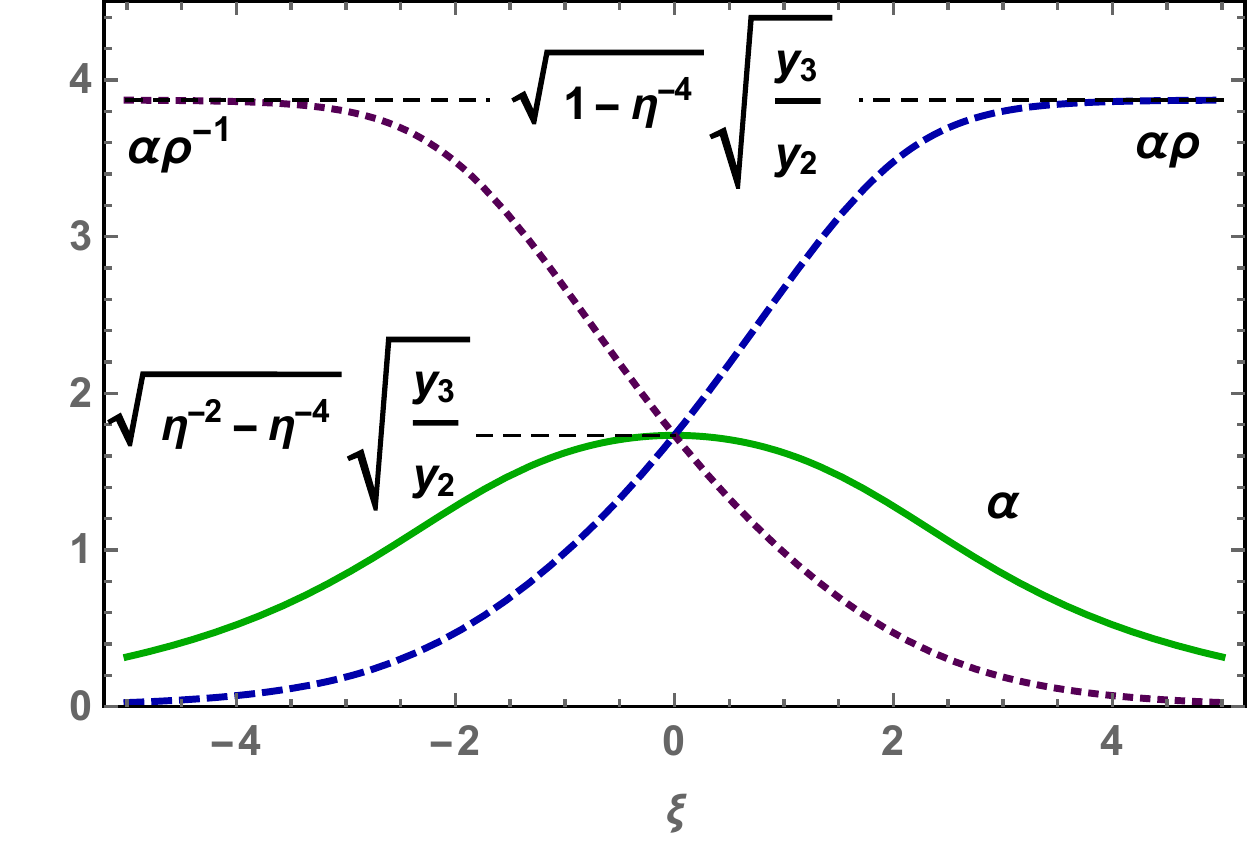}
\caption{The functions $\alpha$ and $\alpha\rho^{\pm1}$ as a function of $\xi$.}
\label{fig:1}
\end{figure}

Let us now consider the matrix $U_L$. 
In all expressions below we leave the implementation of Eq.~(\ref{eq:betasol}) and (\ref{eq:alphasol}) implicit, but in order to get a better feeling for the behavior of the results we show in Fig.~\ref{fig:1} a plot of the relevant functions $\alpha$, and $\alpha\rho^{\pm 1}$ as a function of $\xi$.
Starting with the eigenvector to $y_1$, which is always given by the first column of $\tilde Y^*$ for the regime in Eq.~(\ref{eq:eta2}), one finds
\be
a=\frac{\beta^*}{\eta}\sqrt{\frac{y_1}{y_2}}\,,\qquad \tilde b=-\frac{\eta\alpha\beta^*}{\rho}\sqrt{\frac{y_1}{y_3}}\,.
\ee
Note that $|a|=\eta^{-1} \sqrt{\frac{y_1}{y_2}}$ and $|\tilde b|\leq \eta\sqrt{1-\eta^{-4}}\sqrt\frac{y_1}{y_2}$ (saturated at $\rho=0$).
Moving on to the eigenvector to $y_3$,
note that the first column of $Y$ can never be the longest, whereas the 
third one is the longest when $|Y_{33}|>|Y_{23}|$ or $\alpha\rho\eta^2 \leq \sqrt\frac{y_3}{y_2}$.
In this case ($j=3$) one reads off
\be
b=0\,, \qquad c=\frac{\alpha}{\rho}\eta^2\sqrt{\frac{y_2}{y_3}}\,.
\ee
There appears an upper bound for $c$ given by $|c|\leq \sqrt{\eta^4-1}$, saturated at $\rho=0$.
Finally, when $|Y_{23}|>|Y_{33}|$ or $\alpha\rho\eta^2 \geq  \sqrt\frac{y_3}{y_2}$, the longest column of $Y$ is the second one.
The criterion can only be satisfied for $\eta\geq 2^\frac{1}{4}$, which in turn  means that $j=3$ is guaranteed when this condition is not met.
For $j=2$, the $b$ and $c$ parameters read
\be
b=\frac{\eta}{\alpha\rho}\sqrt{\frac{y_1}{y_3}}\,,\qquad c=\left(\frac{\alpha}{\rho}+\frac{\beta}{\alpha\rho}\right)\eta^2\sqrt{\frac{y_2}{y_3}}\,.
\label{eq:bc2nd}
\ee
Notice that $b$ is bounded by $|b|\leq \eta^3\frac{\sqrt{y_1 y_2}}{y_3}$, which is small but still enhanced compared to the error of the approximation.

We see that as in the Froggatt Nielsen case, some of the rotation angles are suppressed. 
The global bounds (marginalized over $\rho$) are 
\be
|a|^2=\frac{1}{\eta^2}\frac{y_1}{y_2}\,,\qquad |\tilde b|^2\leq \frac{\eta^4-1}{\eta^2}\frac{y_1}{y_2}\,,
\qquad 
|b|^2\leq \eta^6\frac{y_1 y_2}{y_3^2}\,,\qquad |c|^2\leq \eta^4-1\,,
\label{eq:boundsUL}
\ee
which only depend on $\eta$. We notice that $c$ can only be suppressed if $\eta$ is close to one, which in particular implies $j=3$.
We find it interesting that such strong statements as the ones in Eq.~(\ref{eq:boundsUL}) can easily be obtained within our formulation.

The matrix $U_R$ can be obtained by the interchange of $\rho\leftrightarrow \rho^{-1}$ in the above discussion.

To conclude, we stress again that the parametrization chosen here only depends on three free parameters $\rho$, $\eta$, $\arg\beta$, all the while being completely general apart from the mild assumption Eq.~(\ref{eq:eta2}). This has to be contrasted with explicit models, where the Yukawa couplings are typically parametrized in terms of many more parameters, leading to plenty of flat directions when performing fits to the masses and mixings. We therefore hope that our parametrization together with the approximation scheme developed in this paper greatly simplifies the task of finding phenomenologically viable parameters for models with certain textures.

\subsection{Clockwork model}

As a third example we will consider a simple Clockwork \cite{Giudice:2016yja} model similar to the ones of Refs.~\cite{vonGersdorff:2017iym,Patel:2017pct}.
This is an example of a model that does not feature any obvious small expansion parameter, nor do we expect the rotation angles to be small.
This class of models thus nicely illustrates how neither of the two properties are required for our approximation to work.
The Yukawa couplings are generated from the following Clockwork Lagrangian:
\bea
\mathcal L&=&\sum_{i=1}^{N_q}\bar q_i\,\sl p\, q_i-(\bar q^i_R M_qq_L^{i}+\bar q^i_R K_qq_L^{i-1}+h.c.)\nn\\
&&+\sum_{i=1}^{N_u}\bar u_i\,\sl p\,u_i-(\bar u_L^iM_uu_R^{i}+\bar u_L^iK_uu_R^{i-1}+h.c.)\nn\\
&&+\sum_{i=1}^{N_d}\bar d_i\,\sl p\,d_i-(\bar d_L^iM_dd_R^{i}+\bar d_L^iK_dd_R^{i-1}+h.c.)\nn\\
&&+\bar q_L^0\,\sl p\, q_L^0+\bar u_R^0\,\sl  p\,u_R^0+\bar d^0_R\,\sl  p\,d_R^0-(\bar q^0_LH\,d^0_R+\bar d^0_{L}\tilde H\, u^0_{R}+h.c.)\,.
\label{eq:model}
\eea
The last line has a $U(3)$ flavor symmetry that is broken by the matrices $M$ and $K$.
Integrating out the Clockwork fields one finds the effective Lagrangian
\be
\mathcal L_{\rm eff}=\bar q_L^0\,Z_q\,\sl p\, q_L^0+\bar u_R^0\,Z_u\,\sl  p\,u_R^0+\bar d^0_R\,Z_d\,\sl  p\,d_R^0-(\bar q^0_LH\,d^0_R+\bar d^0_{L}\tilde H\, u^0_{R}+h.c.)\,.
\ee
where 
\be
Z_x=\sum_{k=0}^{N_x} (Q_x^\dagger)^k (Q_x)^k\,,\qquad Q_x=M_x^{-1}K_x\,.
\ee 
For $K$ and $M$ random order one matrices, the eigenvalues of the Hermitian matrices $Z_x$ are always greater than one and strongly hierarchical for large $N_x$ \cite{vonGersdorff:2017iym}. 
The physical Yukawa couplings can then be obtained by canonical normalization 
\be
Y_u=Z_q^{-\frac{1}{2}}Z_u^{-\frac{1}{2}}\,,\qquad Y_d=Z_q^{-\frac{1}{2}}Z_d^{-\frac{1}{2}}\,.
\ee
The hierarchical structure of the CKM angles is guaranteed by the common hierarchical factor $Z_q^{-\frac{1}{2}}$.

\begin{figure}
\centering
\includegraphics[width=7cm]{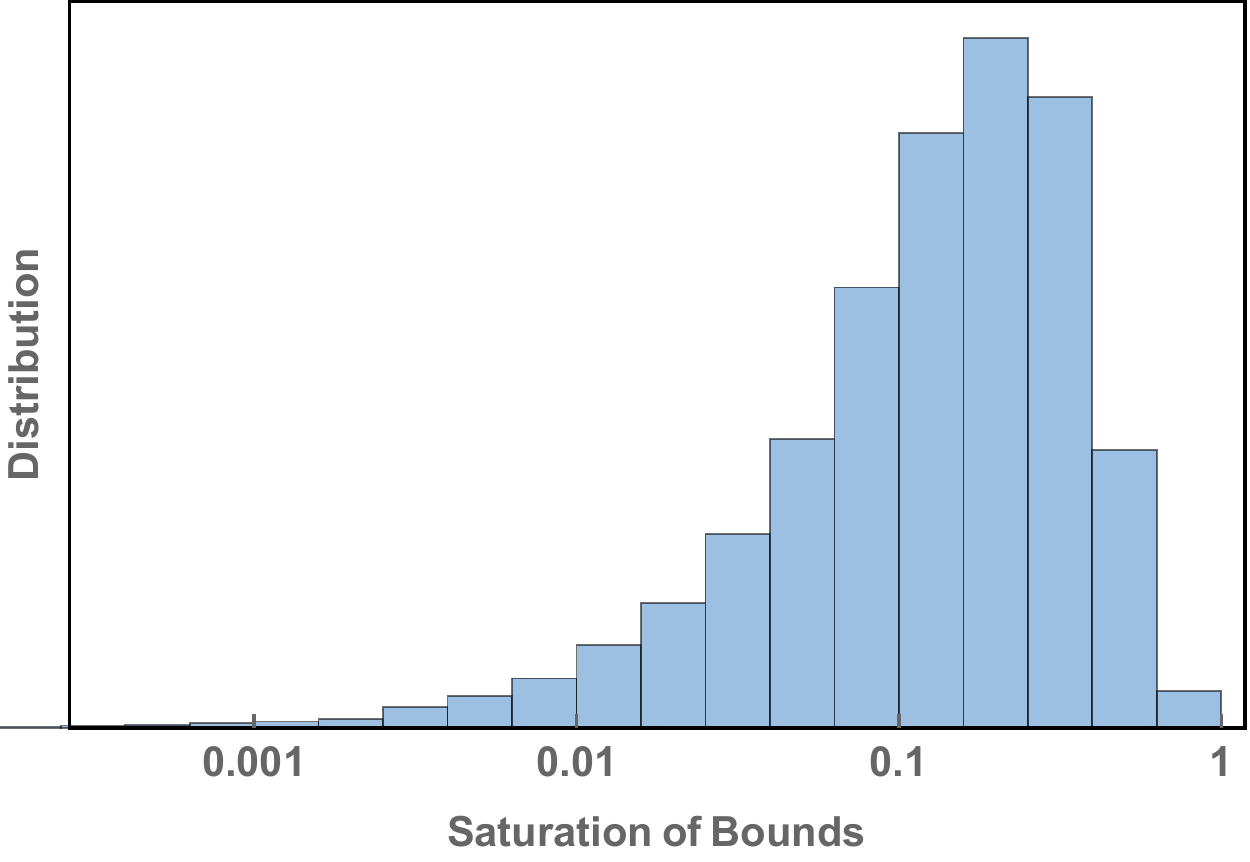}
\caption{Distribution of the quantities $|\delta_{ij}-|V_{ij}|^2|$ normalized to their respective bounds.}
\label{fig:cw}
\end{figure}

We have simulated the Yukawa couplings of this model, using random complex matrices (with flat priors) for $M$ and $K$. We focus on one Yukawa, say $Y_u$ with 
$N_u=N_q=5$ for definiteness.
We have calculated the matrices $U_L$ and $U_R$, using our approximation with $n=\frac{1}{2}$, as well as the error matrices $V_L$ and $V_R$.
The distributions for the $|\delta_{ij}-|V_{ij}|^2|$, normalized to their respective bounds from Eq.~(\ref{eq:errors}), are shown in Fig.~\ref{fig:cw}.~\footnote{We only show the results for $j\neq k$, valid in about 90\% of the simulation.
We use the exact expressions for $X$ and $Z$ given in Eq.~(\ref{eq:A10}) and (\ref{eq:A12}) in order to account for the occasional cases in which the eigenvalues are not too hierarchical. We have also removed cases in which the hierarchies become so mild that the bounds exceed one, which will trivially be satisfied by any unitary matrix.}
These ratios are expected to be smaller than one for our bounds to be correct.
Indeed, no points in our simulation violate the bounds, even though they can come arbitrarily close. 
The median of the distribution is approximately equal to $0.15$.
We have checked that the displayed distribution does not depend significantly neither on the order of the approximation nor on the size of the generated hierarchies. 
In conclusion, this example illustrates that, barring further  information on the structure of the  Yukawa couplings, the global error bounds for our approximation are optimal. 

\section{Conclusions}

Finally let us summarize some of the features of the approximation developed here.
\begin{itemize}
\item
It is {\em universal}, i.e., it does not make any assumptions about the underlying matrix $Y$.
\item
Rather than an expansion scheme (with typically unknown radius of convergence) in which one would need to compute higher and higher terms that need to be summed in order to achieve a desired accuracy, one directly obtains the result up to the desired accuracy in terms of the matrix elements of $Y$.
\item
We have derived a global upper bound for the error matrix $V$. 
Thus, there are no hidden cases where one looses control over the approximation.
One can even obtain an approximate expression for $V$ by plugging in the leading result, see Eq.~(\ref{eq:Vn}). 
\item
In contrast to usual expansion schemes, which  contain often several small parameters, no ambiguities arise as to the relative size of various expansion parameters.
\item
We also have derived two equivalent but nontrivially related approximate expressions for the eigenvalues that can also be taken to arbitrary precision.
\item
The method is well-suited for numerical fits as it is very simple and computationally inexpensive. Moreover, even though the approximation might have large errors away from the physical parameters, in their vicinity (that is, near the "$\chi^2$ minimum") it is necessarily fully under control.
\end{itemize}

\appendix

\section{Error analysis}
\label{sec:error}
In this section we analyze the error of our approximation.
Let $U$ be the matrix whose columns are the eigenvectors of $A$, and let $a_3$ and $a_1$ be the largest and smallest eigenvalues respectively. Furthermore, let the $j$th column of $A^n$ and the  $k$th column of $A^{-n}$ to be the longest.
For now we will assume that $k\neq j$. Furthermore, let $U_{i3}^{(n)}$ be the normalized $j$th column of $A^n$ and $U_{i1}^{(n)}$ be the normalized $k$th column of $A^{-n}$, and let
\be
U_{i2}^{(n)}=\eps_{i\ell m}U_{\ell 3}^{*(n)}U_{m1}^{*(n)}\,.
\ee 
Since $j\neq k$, the matrix $U^{(n)}$ is unitary as $U_{i1}^{(n)}$ and $U_{i3}^{(n)}$ are exactly orthogonal in that case. Let us define the matrix
\be
V=U^\dagger U^{(n)}\,.
\ee
The matrix $V$  is unitary and converges to the identity, encoding the error of our approximation. 
It is easy to obtain explicitly its first and third columns, up to normalization:
\be
V_{i1}
\propto 
a^{-n}_i U^*_{ki}\,,\qquad 
V_{i3}
\propto 
a^{n}_iU^*_{ji}\,.
\label{eq:V0}
\ee
Note that orthogonality of these two columns is exact for $j\neq k$.
In the $abc$ representation $V$ can be given by $K(\Delta a,\Delta b, \Delta c)$ with 
\be
\Delta a=-\frac{U_{k2}}{U_{k1}}\left(\frac{a_1}{a_2}\right)^n\,,\qquad \Delta \tilde b=-\frac{U_{k3}}{U_{k1}}\left(\frac{a_1}{a_3}\right)^n\,,
\label{eq:V1}
\ee
\be
\Delta b=\frac{U_{j1}^*}{U_{j3}^*}\left(\frac{a_1}{a_3}\right)^n\,,\qquad
\Delta c=\frac{U_{j2}^*}{U_{j3}^*}\left(\frac{a_2}{a_3}\right)^n\,.
\label{eq:V2}
\ee
These expressions are exact, but depend on the unknown matrix $U$.
Owing to the longest-column criteria and the unitarity of the matrix $U$ the prefactors can be bounded as follows.
Let $j$ maximize the column of $A^n$ and let $\ell\neq j$. Define 
$\xi_\ell$ to be
 the ratios of the lengths squared of the shorter columns to the longest, $\xi_\ell\equiv (A^{2n})_{\ell\ell}/(A^{2n})_{jj}\leq 1$, or
\be
|U_{j3}|^2a_3^{2n}+|U_{j2}|^2a_2^{2n}+|U_{j1}|^2a_1^{2n}
=\xi_\ell^{-1}
\left( |U_{\ell 3}|^2a_3^{2n}+|U_{\ell 2}|^2a_2^{2n}
+|U_{\ell 1}|^2a_1^{2n}
\right)\,.
\ee
We want to find a lower bound for $|U_{j3}|$ as this quantity appears in the denominators of Eq.~(\ref{eq:V2}).
Maximizing the LHS and minimizing the RHS for fixed $U_{i3}$ gives
\be
|U_{j3}|^2(a_3^{2n}-a_2^{2n})+a_2^{2n}
\geq \xi_\ell^{-1}\left(|U_{\ell 3}|^2(a_3^{2n}-a_1^{2n})+a_1^{2n}\right)\,,
\ee
(which by the way implies that $\xi_\ell\geq (\frac{a_1}{a_3})^{2n}$).
We now use unitarity of the third column $U_{i3}$ to find
\be
|U_{j3}|^2\geq \frac{a_3^{2n}-\xi a_2^{2n}+a_1^{2n} }{(1+\xi)a_3^{2n}-\xi a_2^{2n}+a_1^{2n}}
\ee
where $\xi\equiv \xi_\ell+\xi_{\ell'}$, which is bounded by $2(\frac{a_1}{a_3})^{2n}\leq \xi \leq 2$. 
Finally use unitarity of the $j$th row to get\footnote{Valid only for $a_3^{2n}-\xi a_2^{2n}+a_1^{2n}>0$. We simply define $X_\xi$ to be infinite otherwise.}
\be
\frac{|U_{j1}|^2}{|U_{j3}|^2}+\frac{|U_{j2}|^2}{|U_{j3}|^2}\leq 
\frac{\xi a_3^{2n}-2a_1^{2n} }{a_3^{2n}-\xi a_2^{2n}+a_1^{2n}}
\equiv X_\xi\,.
\ee
The bound is the weakest at $\xi=2$ (when all three columns have equal lengths):
\be
X_\xi\leq  2\frac{a_3^{2n}-a_1^{2n}}{ a_3^{2n}-2 a_2^{2n}+a_1^{2n} }\equiv X\,.
\label{eq:A10}
\ee

Similar reasonings apply to the matrix $A^{-n}$, leading to
\be
\frac{|U_{k3}|^2}{|U_{k1}|^2}+\frac{|U_{k2}|^2}{|U_{k1}|^2}\leq \frac{\zeta a_1^{-2n}-2a_3^{-2n}}{a_1^{-2n}-\zeta a_2^{-2n}+a_3^{-2n}}\equiv Z_\zeta\,,
\ee
where $\zeta$ is the analogous parameter for the inverse matrix, satisfying also $2(\frac{a_1}{a_3})^{2n}\leq \zeta\leq 2$. Again one has the global, $\zeta$-independent bound  \be
Z_\zeta\leq 2\frac{ a_1^{-2n}-a_3^{-2n}}{a_1^{-2n}- 2a_2^{-2n}+a_3^{-2n}}    \equiv Z\,.
\label{eq:A12}
\ee

We can use these results to put upper bounds on the $abc$ parameters of $V$:
\be
|\Delta a|\leq \sqrt Z\left(\frac{a_1}{a_2}\right)^n\,,\qquad |\Delta \tilde b|\leq \sqrt Z\left(\frac{a_1}{a_3}\right)^n
\ee
\be
|\Delta b|\leq \sqrt X\left(\frac{a_1}{a_3}\right)^n\,,\qquad |\Delta c|\leq \sqrt X\left(\frac{a_2}{a_3}\right)^n
\label{eq:six}
\ee
\be
|\Delta b|^2+|\Delta  c|^2\leq X\left(\frac{a_2}{a_3}\right)^{2n}\,,\qquad 
|\Delta a|^2+|\Delta \tilde b|^2\leq Z\left(\frac{a_1}{a_2}\right)^{2n}
\ee
These bounds are exact and only depend on the ratios of eigenvalues. Moreover, for large hierarchies or large $n$  we can  very well approximate  $X\approx  2$, $Z\approx 2$, which are  precisely the bounds one would obtain if it were possible to choose $j$ to maximize $|U_{j3}|$ directly.
On the other hand, one can obtain stronger bounds $X_\xi<X$ and $Z_\zeta< Z$ if one takes the information on the length ratios into account.  


It is now straightforward to put bounds on the matrix $V$:
\bea
\begin{pmatrix}
1-|V_{11}|^2	&	|V_{12}|^2	&	|V_{13}|^2\\
|V_{21}|^2		&	1-|V_{22}|^2&	|V_{23}|^2\\
|V_{31}|^2		&	|V_{32}|^2	&	1-|V_{33}|^2
\end{pmatrix}
&\leq&
\begin{pmatrix}
|\Delta a|^2+|\Delta \tilde b|^2	& 	|\Delta a|^2+|\Delta \tilde b|^2	&	|\Delta b|^2\\
|\Delta a|^2				&	|\Delta a|^2+|\Delta c|^2			&	|\Delta c|^2\\
|\Delta \tilde b|^2		&	|\Delta b|^2+|\Delta c|^2	&	|\Delta b|^2+|\Delta c^2|
\end{pmatrix}\nn\\
&\leq&
\begin{pmatrix}
Z(\frac{a_1}{a_2})^{2n}	& 	Z(\frac{a_1}{a_2})^{2n}	&	X(\frac{a_1}{a_3})^{2n}\\
Z(\frac{a_1}{a_2})^{2n}	&	Z(\frac{a_1}{a_2})^{2n}+X(\frac{a_2}{a_3})^{2n}	&	X(\frac{a_2}{a_3})^{2n}\\
Z(\frac{a_1}{a_3})^{2n}	&	X(\frac{a_2}{a_3})^{2n}	&	X(\frac{a_2}{a_3})^{2n}
\label{eq:bounds}
\end{pmatrix}\,.
\eea
where the first row is correct for any unitary matrix $K(\Delta a,\Delta b,\Delta c)$.
These bounds are exact.
They  are optimized for small values of the ratios $(a_i/a_j)^n$, where $X\approx Z\approx 2$. 
As $V$ is unitary, the bounds become trivially satisfied 
when any of the entries on the RHS of Eq.~(\ref{eq:bounds}) 
become greater than one. This roughly occurs when $(\frac{a_2}{a_3})^{2n}>\frac{1}{4}$ or $(\frac{a_1}{a_2})^{2n}>\frac{1}{4}$.

When it happens that $j=k$, one no longer has that $U_{i1}^{(n)}$ and $U_{i3}^{(n)}$ are exactly orthogonal. In practice, one can simply ignore this small deviation from unitarity (which can be estimated as $U_{i3}^{*(n)} U_{i1}^{(n)}\leq 3(\frac{a_1}{a_3})^n$) but the error estimates are more complicated as $V$ is no longer unitary. Alternatively one can enforce unitarity by computing $\tilde b=b-ac$ instead of extracting it from $U_{i1}^{(n)}$, however, this can introduce large errors in sufficiently pathological cases.

For the purpose of deriving robust upper bounds for the error, we follow a simpler approach and instead  enforce $k\neq j$ (and hence exact unitarity of $U^{(n)}$ and $V$) by taking the second longest column of $A^{-n}$. \footnote{Also, for the simplicity of the argument we will in the following assume that the criteria on the sizes of the columns will coincide with the criteria on the sizes of $|U_{i\ell}|$.}
In this case one obtains a slightly weaker bound as follows. 
As $U_{j3}\geq \sqrt{\frac{1}{3}}$, one has $U_{j1}\leq \sqrt{\frac{2}{3}}$ and hence for the second largest entry one still has a lower bound, $|U_{k1}|\geq\sqrt{\frac{1}{6}}$. Then, unitarity of the $k$th row implies that $|U_{k2}|\leq \sqrt{\frac{5}{6}}$ while unitarity of the third column implies $|U_{k3}|\leq\sqrt\frac{2}{3}$.
As a consequence, one has now that 
\be
|\Delta a|\leq \sqrt{5}\left(\frac{a_1}{a_2}\right)^n\,,\qquad |\Delta \tilde b|\leq 2\left(\frac{a_1}{a_3}\right)^n\,,
\ee
while Eq.~(\ref{eq:six}) still remains valid.
In summary, the errors are now bounded by
\be
\begin{pmatrix}
1-|V_{11}|^2	&	|V_{12}|^2	&	|V_{13}|^2\\
|V_{21}|^2		&	1-|V_{22}|^2&	|V_{23}|^2\\
|V_{31}|^2		&	|V_{32}|^2	&	1-|V_{33}|^2
\end{pmatrix}
\leq
\begin{pmatrix}
 5(\frac{a_1}{a_2})^{2n}	&  5	(\frac{a_1}{a_2})^{2n}	&	 2(\frac{a_1}{a_3})^{2n}\\
 5(\frac{a_1}{a_2})^{2n}	&	5(\frac{a_1}{a_2})^{2n}+2(\frac{a_2}{a_3})^{2n}		& 2(\frac{a_2}{a_3})^{2n}\\
4(\frac{a_1}{a_3})^{2n}	& 2	(\frac{a_2}{a_3})^{2n}	& 2	(\frac{a_2}{a_3})^{2n}
\end{pmatrix}\,.
\label{eq:bounds2}
\ee
It is clear that one could also work with the case of the longest column of $A^{-n}$ and the second longest column of $A^n$, but we will not spell out this case in detail.

For $n=\frac{1}{2}$, one replaces Eq.~(\ref{eq:V0}) by
\be
(V_L)_{i1}\propto
y_i^{-1} (U_R)^*_{ki}\,,\qquad
(V_L)_{i3}\propto y_i (U_R)^*_{ji}\,,
\ee
and
\be
(V_R)_{i1}\propto
y_i^{-1}(U_L)^*_{ki}\,,\qquad
(V_R)_{i3}\propto
y_i(U_L)^*_{ji}\,.
\ee
The bounds one obtains are then given by setting $a_i=y_i^2$ and $n=\frac{1}{2}$ (or $a_i^n\to y_i$) in the previous expressions.

It should be noted that the error matrix $V$ can itself be computed approximatively by replacing in Eqns.~(\ref{eq:V1}) and (\ref{eq:V2}) the $n$th order result for $U$. 
This can be  useful as sometimes the quantities $\Delta a$, $\Delta b$, $\Delta c$ are more suppressed than the conservative upper bounds presented here, leading to much more accurate results. 
Calling this matrix $V^{(n)}$, it is given explicitly by
\be
V^{(n)}
\approx
\begin{pmatrix}
1	& -\frac{U^{(n)}_{k2}}{U^{(n)}_{k1}}(\frac{a_1}{a_2} )^n	& \frac{U^{*(n)}_{j1}}{U^{*(n)}_{j3}}(\frac{a_1}{a_3})^n\\
\frac{U^{*(n)}_{k2}}{U^{*(n)}_{k1}}(\frac{a_1}{a_2} )^n	&1					&\frac{U_{j2}^{*(n)}}{U_{j3}^{*(n)}}(\frac{a_2}{a_3})^n\\
\frac{U^{*(n)}_{k3}}{U^{*(n)}_{k1}}(\frac{a_1}{a_3})^n&-\frac{U^{(n)}_{j2}}{U^{(n)}_{j3}}(\frac{a_2}{a_3})^n&1
\end{pmatrix}\,,
\label{eq:Vn}
\ee 
where, in addition to replacing $U$ by  $U^{(n)}$ we have expanded in the hierarchies to leading order.
Notice for the case of half-integer $n$, $V_L$ depends on $U_R$ and vice versa.


\bibliography{paper}

\begin{thebibliography}{10}

\bibitem{Antusch:2013jca}
S.~Antusch and V.~Maurer, ``{Running quark and lepton parameters at various
  scales},'' {\em JHEP}, vol.~11, p.~115, 2013, 1306.6879.

\bibitem{Froggatt:1978nt}
C.~D. Froggatt and H.~B. Nielsen, ``{Hierarchy of Quark Masses, Cabibbo Angles
  and CP Violation},'' {\em Nucl. Phys.}, vol.~B147, pp.~277--298, 1979.

\bibitem{Haba:2000be}
N.~Haba and H.~Murayama, ``{Anarchy and hierarchy},'' {\em Phys. Rev.},
  vol.~D63, p.~053010, 2001, hep-ph/0009174.

\bibitem{Cabrer:2011qb}
J.~A. Cabrer, G.~von Gersdorff, and M.~Quiros, ``{Flavor Phenomenology in
  General 5D Warped Spaces},'' {\em JHEP}, vol.~01, p.~033, 2012, 1110.3324.

\bibitem{Dudas:2013pja}
E.~Dudas, G.~von Gersdorff, S.~Pokorski, and R.~Ziegler, ``{Linking Natural
  Supersymmetry to Flavour Physics},'' {\em JHEP}, vol.~01, p.~117, 2014,
  1308.1090.

\bibitem{Gherghetta:2000qt}
T.~Gherghetta and A.~Pomarol, ``{Bulk fields and supersymmetry in a slice of
  AdS},'' {\em Nucl. Phys.}, vol.~B586, pp.~141--162, 2000, hep-ph/0003129.

\bibitem{Huber:2000ie}
S.~J. Huber and Q.~Shafi, ``{Fermion masses, mixings and proton decay in a
  Randall-Sundrum model},'' {\em Phys. Lett.}, vol.~B498, pp.~256--262, 2001,
  hep-ph/0010195.

\bibitem{Alonso:2018bcg}
R.~Alonso, A.~Carmona, B.~M. Dillon, J.~F. Kamenik, J.~Martin~Camalich, and
  J.~Zupan, ``{A clockwork solution to the flavor puzzle},'' {\em JHEP},
  vol.~10, p.~099, 2018, 1807.09792.

\bibitem{Babu:2009fd}
K.~S. Babu, ``{TASI Lectures on Flavor Physics},'' in {\em {Proceedings of
  Theoretical Advanced Study Institute in Elementary Particle Physics on The
  dawn of the LHC era (TASI 2008): Boulder, USA, June 2-27, 2008}},
  pp.~49--123, 2010, 0910.2948.

\bibitem{Barbieri:1995uv}
R.~Barbieri, G.~R. Dvali, and L.~J. Hall, ``{Predictions from a U(2) flavor
  symmetry in supersymmetric theories},'' {\em Phys. Lett.}, vol.~B377,
  pp.~76--82, 1996, hep-ph/9512388.

\bibitem{Barbieri:1997tu}
R.~Barbieri, L.~J. Hall, and A.~Romanino, ``{Consequences of a U(2) flavor
  symmetry},'' {\em Phys. Lett.}, vol.~B401, pp.~47--53, 1997, hep-ph/9702315.

\bibitem{Giudice:2016yja}
G.~F. Giudice and M.~McCullough, ``{A Clockwork Theory},'' {\em JHEP}, vol.~02,
  p.~036, 2017, 1610.07962.

\bibitem{vonGersdorff:2017iym}
G.~von Gersdorff, ``{Natural Fermion Hierarchies from Random Yukawa
  Couplings},'' {\em JHEP}, vol.~09, p.~094, 2017, 1705.05430.

\bibitem{Patel:2017pct}
K.~M. Patel, ``{Clockwork mechanism for flavor hierarchies},'' {\em Phys.
  Rev.}, vol.~D96, no.~11, p.~115013, 2017, 1711.05393.

\end{thebibliography}
\bibliographystyle{hieeetr}

\end{document}